\documentclass[10pt]{article} 
\usepackage{graphicx,amssymb,amsmath,epsf,rotate,amsfonts,latexsym} 
\usepackage{xcolor}

\begin{document} 
\begin{flushright} 
SINP/TNP/2011/12, ~DO-TH-11/20
\end{flushright} 
 
\vskip 30pt 
 
\begin{center}
  {\Large \bf $R$-Parity violating flavor symmetries, recent neutrino
    data and absolute neutrino mass scale}
 \\
  \vspace*{1cm} 
\renewcommand{\thefootnote}{\fnsymbol{footnote}} 
{ {\sf Gautam Bhattacharyya${}^1$}, {\sf Heinrich P\"as${}^{2}$}, 
 and {\sf Daniel Pidt${}^{2}$}} \\
  \vspace{10pt} 
   {\small ${}^{1)}$ {\em Saha Institute of Nuclear Physics,
      1/AF Bidhan Nagar, Kolkata 700064, India} \\
    ${}^{2)}$ {\em Fakult\"at f\"ur Physik, Technische Universit\"at Dortmund,
      D-44221, Dortmund, Germany} 
}
\normalsize
\end{center} 

\begin{abstract} 
\noindent
We study the r\^ole of a very general type of flavor symmetry in
controlling the strength of $R$-parity violation in supersymmetric
models.  We assume that only leptons are charged under a global
symmetry whose breaking induces lepton number (and, hence, $R$-parity)
violation. The charge assignments of leptons under this symmetry are
such that the total number of independent lepton number violating
couplings is reduced from 39 to 6.  The most severe constraints on
these flavor-correlated couplings arise from neutrino masses and
mixing as well as from the non-observation of $K_L \to \mu e$.  We
find that such a scenario predicts an almost vanishing smallest
neutrino mass eigenvalue, allowing the upcoming generation of
neutrinoless double beta decay experiments to shed light on the
hierarchy.
\end{abstract} 

\renewcommand{\thesection}{\Roman{section}} 
\setcounter{footnote}{0} 
\renewcommand{\thefootnote}{\arabic{footnote}}


\section{Introduction}
$R$-parity in supersymmetry is a discrete symmetry which is defined as
$R_p = (-1)^{3B+L+2S}$, where $B$, $L$ and $S$ are the baryon number,
lepton number and spin of the particle respectively (see
\cite{rpar,spon_rpv,reviews,Kao:2009fg}). All standard model (SM)
particles have $R_p =1$, while all superparticles have $R_p = -1$. The
assumption of $R$-parity conservation in supersymmetric models is
quite {\em ad hoc}, as this is not supported by any deep underlying
principle.  Historically, it was imposed to keep the proton
stable. However, proton decay requires a simultaneous presence of $B$
and $L$ violation.  Therefore, dropping all $R_p$ violating (RpV)
couplings in one go is certainly an overkill \cite{pdecay}. Still, in
conventional supersymmetric theories $R_p$ conservation is imposed
primarily for the sake of convenience, as otherwise the number of
independent parameters in the minimal supersymmetric standard model
(MSSM), which is already very large and difficult to handle, is
augmented by a set of new RpV parameters.  Moreover, conserved $R_p$
implies that the lightest supersymmetric particle is stable, which
leads to a plausible dark matter candidate, and also attributes
supersymmetry with a characteristic missing energy signature in
colliders. On the other hand, if lepton number is violated, one
distinct advantage is that neutrino masses can be generated via a
perfectly renormalizable interaction
\cite{numass,Abada:2001zh,Grossman:2003gq} without the need of
introducing any right-handed neutrinos.

The most general superpotential with explicit RpV couplings is given
by \cite{reviews}
\begin{align}
  W_{RPV}=\mu_iL_iH_u+\frac{1}{2}\lambda_{ijk}L_iL_jE^C_k+
\lambda'_{ijk}L_iQ_jD^C_k+\frac{1}{2}\lambda''_{ijk}U^C_iD^C_jD^C_k,
\label{eqn:WRPV}
\end{align}
where $i$, $j$ and $k$ are the three quark and lepton generation
indices. Here, $Q_i$ and $L_i$ are SU(2)-doublet quark and lepton
superfields, respectively; $D^C_i$ and $E^C_i$ are SU(2)-singlet
superfields for down-type quarks and charged leptons, respectively;
and, $H_u$ is the Higgs superfield that generates the mass of the
up-type quarks.  This introduces 48 new couplings: 3 $\mu_i$-type, 9
$\lambda_{ijk}$-type (note the antisymmetry in the first two indices),
27 $\lambda'_{ijk}$-type and 9 $\lambda''_{ijk}$-type couplings (note
the antisymmetry in the last two indices). Only the $\lambda''$
couplings are $B$ violating, the rest are all $L$ violating.  Besides,
new RpV soft terms appear which introduce more unknown parameters.  We
do not explicitly write down these soft terms but will mention about
the relevant ones in appropriate places. Dealing with so many new
parameters substantially reduces the predictivity of the model.  At
this point there are two ways to proceed. One may either take one or
two RpV couplings as non-vanishing at a time and study their
implications, or, apply some suitable flavor symmetry to relate one
coupling to another \cite{Banks:1995by,Bhattacharyya:1998vw}.  We
shall take the latter approach in this paper (see also
\cite{Dreiner:2007uj} which does not differentiate between lepton
doublet and down-type Higgs superfields for a different approach
following a similar spirit).  We show that with a simple flavor
hypothesis we can bring down the number of totally independent RpV
(more specifically, $L$ violating) couplings to only six. These
couplings induce neutrino Majorana masses, and if the neutrino mixing
matrix is tri-bimaximal (TBM) then the number of independent $L$
violating couplings can be further reduced to four, a scenario which
prefers an inverse neutrino mass hierarchy.

\section{A generic flavor model}
We assume that the Yukawa structure leading to the masses and mixing
of quarks and charged leptons is fixed by some unspecified global
symmetry. This symmetry also ensures baryon number conservation.
There is a second global symmetry ($X$), an abelian horizontal
symmetry, which is at the centre of our attention. {\em Only} leptons
are charged under $X$, such that for each generation $i$,
\begin{eqnarray} 
Q_X(L_i) = - Q_X(E^C_i) \, . 
\label{eqn:Bcharges}
\end{eqnarray} 
We assume that the $Q_X$ charges of different generations are all
positive. The horizontal symmetry is explicitly broken by a small
parameter $\varepsilon < 1$, whose charge under $X$ is
$Q_X(\varepsilon) = -1$. If the total charge of a given superpotential
term is $n$, then the term is suppressed by $\varepsilon^n$.  As an
example, if $Q = Z_N$, then the suppression would be
$\varepsilon^{n(\text{mod}N)}$ \cite{Banks:1995by}.

Now we look at the consequences of Eq.~(\ref{eqn:Bcharges}) for the 48
RpV couplings of Eq.~(\ref{eqn:WRPV}). Since $B$-number is conserved,
all the $\lambda''$ couplings vanish right away. Since only leptons
are charged under $X$, it follows that $Q_X(L_iQ_jD^C_k) = Q_X
(L_iH_u) = Q_X(L_i)$, and hence $\lambda'_i \equiv \lambda'_{ijk}
\simeq \tilde\mu_i \equiv \mu_i/\mu$, where the supersymmetry
preserving $\mu$ parameter is assumed to be of the same order as the
supersymmetry breaking soft masses ($\tilde m$). Turning our attention
to the $L_iL_jE^C_k$ operator, we notice that when $j=k$, the same
argument as above leads to $\lambda_{ijj} \simeq \lambda'_{ijk} \simeq
\tilde\mu_i$. Thus 39 {\em a priory} independent $L$-violating
couplings basically boil down to only 6:
\begin{eqnarray} 
\label{six_coup}
\tilde\mu_i \left(\simeq \lambda'_{ijk} \simeq \lambda_{ijj}\right) \, ,~~
\lambda_{123}, ~\lambda_{132}, ~\lambda_{231} \, .    
\end{eqnarray} 
Thanks to the flavor symmetry, the $L$-violating bilinear soft
parameters $B_i$ would be aligned to the corresponding superpotential
parameters $\mu_i$ as well, i.e. $\tilde B_i \equiv B_i/\tilde{m}^2
\simeq \tilde\mu_i$. It should be noted that when we say that two
couplings are related, we mean that they have a common suppression
factor $\epsilon^{Q_X}$. Indeed, there are order-one uncertainties in
the actual coefficients of the operators, for which reason we have
used a `near-equality' sign in Eq.~(\ref{six_coup}).  Now we come to
the relative sizes of the $L$-violating couplings. The suppression
would depend on the sum of $Q_X$ charges of the associated lepton
fields as a power of $\varepsilon$. More specifically,
\begin{eqnarray} 
\label{suppression} 
\tilde\mu_{i} \simeq \tilde B_i \simeq \lambda'_i \sim  \varepsilon^{Q_X(L_{i})}, ~~
\lambda_{ijk} \sim \varepsilon^{\left\{Q_X(L_{i})+Q_X(L_{j})+Q_X(E_{k}^{C})\right\}} \, . 
\end{eqnarray}  
Eventually, we shall provide a specific demonstration with $Q =
Z_{N_1} \times Z_{N_2}$ \cite{Banks:1995by}, which means there are all
together 6 charges for the three lepton generations.

\begin{table}
	\centering
	\begin{tabular}{cccc}
Our couplings & Related to & Existing limits (Sources) & Refs. \\ \hline
$\tilde\mu_{i}$ & $\tilde\mu_{i}$, $\lambda'_{ijk}$, $\lambda_{ijj}$ &
$1.5\times 10^{-6} \left(M_{\tilde \chi}\right)\ [m_{\nu_{i}}]$&\cite{Abada:2001zh} 
\\ 
$\lambda_{123}$ & $\lambda_{123}$ & $0.03\,\left(\tilde{\tau}_R\right)\ [V_{ud}]$ &
\cite{Kao:2009fg} 
\\ 
$\lambda_{132}$ & $\lambda_{132}$ & $0.03\,\left(\tilde{\mu}_R\right)\ 
[R_\tau]$ & \cite{Kao:2009fg}
\\ 
$\lambda_{231}$ & $\lambda_{231}$ &
$0.05\,\left(\tilde{e}_R\right)\ [R_\tau]$ & \cite{Kao:2009fg} 
\\
\end{tabular}
\caption[]{\small The list of the six independent couplings and the standard
  couplings they are related to by the flavor symmetry $X$. The three
  $\tilde\mu_i$ couplings are of the same order of magnitude as 36 out of 39
  {\em a priori} independent RpV couplings.  A mass of 100 GeV is
  assumed for the superparticles exchanged in the processes
  involved. These superparticles are indicated within first bracket
  right after the bounds (the weak gaugino mass $M_\chi$ and the three
  scalar leptons $\tilde \ell_R$).  The entries in the square brackets
  specify the different observables from which origin the bounds
  originate. Here, 
$R_{\tau}=\Gamma(\tau^{-}\rightarrow\mu^{-}\overline{\nu}_{\mu}\nu_{\tau})/
\Gamma(\mu^{-}\rightarrow e^{-}\overline{\nu}_{e}\nu_{\mu})$.}
	\label{tab:CouplingDependencies}
\end{table}

Many RpV couplings which are not so strongly constrained may now be
related by Eq.~(\ref{suppression}) to the ones which are severely
constrained by experiments.  The existing bounds on the individual and
product couplings can be found in the reviews \cite{reviews}.

\section{Neutrino masses and mixing}
One of the high points of $R$-parity violation is that it generates
neutrino masses and mixing through a perfectly renormalizable
interaction without the need of introducing any right-handed neutrino.
This has already been studied at various levels of detail
\cite{numass,Abada:2001zh,Grossman:2003gq}. In this work we will
follow the notation of \cite{Grossman:2003gq}. The neutrino masses, in
the basis in which all the sneutrino vacuum expectation values vanish,
can be written as
\begin{align}
m_{ij} &\approx
\frac{\cos^{2}\beta}{\tilde{m}}\mu_{i}\mu_{j}+\frac{g^{2}}
{64\pi^{2}\cos^{2}\beta}\frac{B_{i}B_{j}}{\tilde{m}^{3}}+
\frac{g^{2}}{64\pi^{2}\cos\beta}\frac{\mu_{i}B_{j}+\mu_{j}B_{i}}{\tilde{m}^{2}}
 \nonumber \\ &+
\sum_{k}\frac{3}{16\pi^{2}}gm_{d_{k}}\frac{\mu_{i}\lambda'_{jkk}+\mu_{j}\lambda'_{ikk}}
{\tilde{m}}
 +
 \sum_{k}\frac{1}{16\pi^{2}}gm_{e_{k}}\frac{\mu_{i}\lambda_{jkk}+\mu_{j}\lambda_{ikk}}
{\tilde{m}} \label{eqn:nmmgeneral}
        \\ &+
        \sum_{l,k}\frac{3}{8\pi^{2}}\lambda'_{ilk}\lambda'_{jkl}\frac{m_{d_{l}}m_{d_{k}}}
            {\tilde{m}^2_q} \mu \tan\beta +
            \sum_{l,k}\frac{1}{8\pi^{2}}\lambda_{ilk}\lambda_{jkl}\frac{m_{e_{l}}m_{e_{k}}}
                {\tilde{m}^2} \mu \tan\beta, \nonumber
\end{align}
where $m_{d_{i}}$ ($m_{e_{i}}$) denote the masses of the down-type
quarks (charged leptons).  A comment on the approximations made above
is in order. We have denoted the squark masses by $\tilde{m}_q$ and
assumed them to be somewhat heavier than a common mass scale $\tilde
m$ assumed for the sleptons and weak gauginos/Higgsinos. This
approximation may be crude but is good enough for our
order-of-magnitude estimate of the RpV couplings.  In
Eq. (\ref{eqn:nmmgeneral}), the first line accounts for the tree level
and one loop contributions from bilinear couplings only, the second
line represents the one loop contributions from both bilinear and
trilinear couplings, while the last line stands for one loop
contributions from trilinear couplings only. The possibility of large
left-right squark/slepton mixing which may be induced by large
$\tan\beta (\equiv v_u/v_d)$ has been taken into account in the purely
trilinear loop dynamics.  The tree level $\mu_i\mu_j$ contribution
generates a rank-one mass matrix and therefore yields only one mass
eigenvalue. Since, in our case, $B_i$, $\lambda'_i$, $\lambda_{ijj}$
are all proportional to $\mu_i$, even after including their
contributions the rank-one nature of the mass matrix does not change.
What breaks the alignment and yields more non-vanishing eigenvalues is
the contribution from the purely trilinear loops involving
$\lambda_{ijk} (i\ne j \ne k)$, since these couplings are not aligned
with $\mu_i$.

This leaves us with the remaining three couplings, namely,
$\lambda_{123}$, $\lambda_{132}$ and $\lambda_{231}$, no two indices
of which are the same, for generating the second {\em mandatory} and
the third {\em optional} nonvanishing neutrino masses and the three
mixing angles (two large and one small). Note that the existing bounds
on $\lambda_{ijk}$ with $i\ne j \ne k$ are relatively less stringent
-- see Table \ref{tab:CouplingDependencies}.

Different low energy processes, especially some lepton flavor
violating decays, yield important constraints on trilinear product
couplings \cite{Bhattacharyya:2009hb,Dreiner:2006gu,Dreiner:2001kc}.
Due to the smallness of {\em most} of the couplings as shown in the
first row of Table \ref{tab:CouplingDependencies}, these constraints
are in almost all cases easily satisfied. The bounds emerging from the
nonobservation of $K_{L}^{0}\rightarrow\mu \overline{e}/e
\overline{\mu}$ \cite{Dreiner:2006gu,Dreiner:2001kc}, namely,
\begin{align}
	\lambda_{ijk}\lambda'_{lmn}	&<	
6.7\times10^{-9} ~m_{\tilde{\nu}_{L3}}^{2}~/~(100\,\text{GeV})^{2} \, , 
	\label{prod-1}
\end{align}
with the combinations $(ijk)(lmn) : (312)(312), (312)(321),
(321)(312), (321)(321)$, play a crucial r\^ole in neutrino mass/mixing
model building in our scenario, as we shall see later.  Due to the
specific inter-connections among RpV couplings owing to the flavor
symmetry, Eq.~(\ref{prod-1}) leads to the following limits:
\begin{align}
	\lambda_{132}\lambda'_{3} \, , ~~ \lambda_{231}\lambda'_{3}
        &<	
6.7\times10^{-9} ~m_{\tilde{\nu}_{L3}}^{2}~/~(100\,\text{GeV})^{2} \, . 
	\label{prod-2}
\end{align}

If we set the numerical values of the couplings near their upper limits (see
Table \ref{tab:CouplingDependencies}), they turn out to be large
enough to offset the loop suppression factors. The mass matrix entries
can then be written with only six RpV couplings as
\begin{align}
m_{ee} &\approx	a\mu_{1}\mu_{1} + 
\frac{1}{8\pi^{2}}\lambda_{123}\lambda_{132}\frac{m_{\mu}m_{\tau}}{\tilde{m}^{2}}\mu
\tan\beta,\nonumber\\
m_{e\mu} &\approx a\mu_{1}\mu_{2}+
\frac{1}{8\pi^{2}}\lambda_{123}\lambda_{232}
\frac{m_{\tau}m_{\mu}}{\tilde{m}^{2}}\mu\tan\beta+\frac{1}{8\pi^{2}}
\lambda_{213}\lambda_{131}\frac{m_{\tau}m_{e}}{\tilde{m}^{2}}\mu\tan\beta,\nonumber\\
m_{e\tau} &\approx
a\mu_{1}\mu_{3}+\frac{1}{8\pi^{2}}\lambda_{132}\lambda_{323}
\frac{m_{\mu}m_{\tau}}{\tilde{m}^{2}}\mu\tan\beta+\frac{1}{8\pi^{2}}\lambda_{312}
\lambda_{121}
\frac{m_{\mu}m_{e}}{\tilde{m}^{2}}\mu\tan\beta,
\label{eqn:nmmDominant}	\\
m_{\mu\mu} &\approx a\mu_{2}\mu_{2}+
\frac{1}{8\pi^{2}}\lambda_{231}\lambda_{213}
\frac{m_{e}m_{\tau}}{\tilde{m}^{2}}\mu\tan\beta,\nonumber	\\
m_{\mu\tau} &\approx
a\mu_{2}\mu_{3}+\frac{1}{8\pi^{2}}\lambda_{231}\lambda_{313}
\frac{m_{\tau}m_{e}}{\tilde{m}^{2}}\mu\tan\beta+
\frac{1}{8\pi^{2}}\lambda_{321}\lambda_{212}\frac{m_{\mu}m_{e}}{\tilde{m}^{2}}\mu\tan\beta,
\nonumber\\
m_{\tau\tau} &\approx a\mu_{3}\mu_{3}+
\frac{1}{8\pi^{2}}\lambda_{312}\lambda_{321}
\frac{m_{\mu}m_{e}}{\tilde{m}^{2}}\mu\tan\beta, \nonumber 	
\end{align}
with
\begin{align}
a &= \frac{\cos^{2}\beta}{\tilde{m}}+\sum_{k}\frac{3gm_{d_{k}}}
{8\pi^{2}\tilde{m}^{2}}+\sum_{k}\frac{gm_{e_{k}}}{8\pi^{2}\tilde{m}^{2}}
+ \sum_{k,l}\frac{3m_{d_{l}}m_{d_{k}}}{8\pi^{2}\mu\tilde{m}_{q}^{2}}\tan\beta \, .
\label{eqn:PropFactor}
\end{align}
With this mass matrix we try to reproduce the neutrino oscillation
data, namely, the two mass-squared differences ($\Delta m_{21}^{2}$
and $\Delta m_{31}^{2}$) and the three mixing angles ($\theta_{12}$,
$\theta_{23}$ and $\theta_{13}$). For simplicity we assume that all
the phases in the neutrino mixing matrix are zero.  Since neutrino
oscillation analysis probes only the mass-squared differences and not
their absolute values, we need to assume the hierarchy
(normal/inverted) of the masses and the size of the smallest
eigenvalue to fix the other two masses. There is no lower limit on the
smallest neutrino mass eigenvalue, it can still be zero.

We take the best fit values of the neutrino mass-squared differences
from \cite{GonzalezGarcia:2010er}: $\Delta m^2_{21} = 7.59 \times
10^{-5} ~\text{eV}^2$, $\Delta m^2_{31} (\text{IH})= - 2.40 \times
10^{-3} ~\text{eV}^2$, $\Delta m^2_{31} (\text{NH})= 2.51 \times
10^{-3} ~\text{eV}^2$.  Above, NH stands for `normal hierarchy' and IH
stands for `inverted hierarchy' of neutrino masses. The two mixing
angles $\theta_{12}$ and $\theta_{23}$ are set to their TBM values
(using best fit values instead does not lead to significant
changes). Very recently, two long-baseline accelerator experiments T2K
and MINOS, both probing $\nu_\mu \to \nu_e$ appearance, have reported,
for the first time, a non-zero measurement of $\theta_{13}$. T2K has
observed 6 electron-like events against an estimated background of
1.5, thus discarding $\theta_{13} = 0$ at the level of 2.5$\sigma$
\cite{Abe:2011sj}. The MINOS experiment observes 62 electron-like
events against an expected 49, thus disfavoring $\theta_{13} = 0$ at
1.5$\sigma$ \cite{minos}. A new global fit suggests $\sin^2
\theta_{13} = 0.021 (0.025) \pm 0.007$ with old (new) reactor fluxes
\cite{Fogli:2011qn}. The central value corresponds to $\theta_{13}
\approx 9^\circ$. A second global fit can be found in
\cite{Schwetz:2011zk}. Both indicate $\theta_{13}>0$ with a
significance of about 3$\sigma$.  We shall see below that in our
scenario whether $\theta_{13}$ is vanishing or non-vanishing plays an
important r\^ole in predicting whether neutrino mass hierarchy is {\em
  normal} ($\Delta m^2 \equiv m_3^2 - 0.5(m_2^2 + m_1^2) > 0$) or {\em
  inverted} ($\Delta m^2 < 0$). Since none of the two experiments has
so far conclusively established a nonzero value of $\theta_{13}$, we
take both the paradigms, namely, $\theta_{13} = 0$ and $\theta_{13}
\neq 0$, and study what do they imply on the choices of RpV parameters
and whether we can predict the nature of mass hierarchy.

\section{Tri-bimaximal mixing}
The TBM structure immediately implies that
$\left|m_{e\mu}\right| = \left|m_{e\tau}\right|$ and
$\left|m_{\mu\mu}\right| = \left|m_{\tau\tau}\right|$, regardless of
whether the lightest mass eigenvalue is vanishing or not, and also
irrespective of whether the neutrino mass hierarchy is normal or
inverted.  For our couplings this can be comfortably realized by
setting $|\lambda_{123}|=|\lambda_{132}|$ and $|\mu_{2}|=|\mu_{3}|$ --
see Eq.~(\ref{eqn:nmmDominant}).  This means that we can parametrize
the mass matrix with four independent RpV parameters instead of six,
which of course improves the predictivity of the model. Dropping the
terms in the loop contribution proportional to the electron mass, we
obtain
\begin{eqnarray}
  m_{e\mu} \approx m_{e\tau} \approx 
  a\mu_{1}\mu_{2}-\frac{1}{8\pi^{2}}\lambda_{123}\mu_{3}
\frac{m_{\tau}m_{\mu}}{\tilde{m}^{2}}\tan\beta \, . 
\label{approx-emutau}
\end{eqnarray}
Clearly, under this situation, the absolute values for the tree-level
contributions to $m_{\mu\mu}\sim a\mu_2\mu_2$, $m_{\mu\tau}\sim
a\mu_2\mu_3$ and $m_{\tau\tau}\sim a\mu_3\mu_3$ are the same.  Setting
all CP-violating phases to zero, the TBM mixing matrix takes the form
\cite{Harrison:2002er} 
\begin{align}
U_{\text{TBM}} &= \begin{pmatrix}
\sqrt{\frac{2}{3}}	& \frac{1}{\sqrt{3}}	&	0	\\
-\frac{1}{\sqrt{6}}	& \frac{1}{\sqrt{3}} & \frac{1}{\sqrt{2}} \\
\frac{1}{\sqrt{6}}	& -\frac{1}{\sqrt{3}}	&	\frac{1}{\sqrt{2}}
\end{pmatrix}.
\label{eqn:UTBM}
\end{align}
To fix the numerical values of the mass matrix from
$m=U^T_{\text{TBM}} \times \text{diag}(m_{1},m_{2},m_{3}) \times
U_{\text{TBM}}$, all we need to decide is the mass hierarchy (normal
or inverted) and the smallest mass eigenvalue.

{\em Inverted hierarchy:}~ We first consider the case of inverted
hierarchy with $m_{3}=0$.  This choice additionally demands
$m_{\mu\tau}=-m_{\mu\mu}$.  One obtains
\begin{align}
m = \begin{pmatrix}
4.92\times 10^{-2} & 2.56 \times 10^{-4}	& -2.56 \times 10^{-4}	\\
2.56 \times 10^{-4} & 2.47 \times 10^{-2} & -2.47 \times 10^{-2}	\\
-2.56 \times 10^{-4} & -2.47 \times 10^{-2} & 2.47 \times 10^{-2}	
\end{pmatrix}\,\text{eV}\ \text{(IH, TBM, $m_3=0$)}.
\label{ih-tbm-m30}
\end{align}
By setting $\mu_2 = - \mu_3$ and keeping $\lambda_{231} \lesssim
\lambda_{123}$, we obtain a rough analytical solution using
Eqs. (\ref{suppression}) and (\ref{eqn:nmmDominant}):
\begin{eqnarray}
\left|\mu_{2}\right| =  \left|\mu_{3}\right| \approx	
\sqrt{a^{-1}  m_{\mu\mu}}, ~~~
\lambda_{123} &\approx
\sqrt{\displaystyle\frac{4\pi^{2}m_{ee}\tilde{m}^{2}}
{m_{\mu}m_{\tau}\mu\tan\beta}}, ~~~
\mu_{1}	\approx 
\displaystyle\frac{m_{e\mu}+\mu_{2}\lambda_{123} m_{\tau}m_{\mu}\tan\beta/ 
\left(8\pi^{2}\tilde{m}^{2}\right)}{\left(a \mu_2 \right)}.	
\label{eqn:Mu2}
\end{eqnarray}
Putting $\tilde{m} = \mu =100\,\text{GeV}$ and
$\tilde{m}_{q}=300\,\text{GeV}$ in Eq.~(\ref{eqn:Mu2}) we obtain a
solution (with $\mu_2 = -\mu_3$, and $\lambda_{132} = - \lambda_{123}$) 
\begin{eqnarray}
  \tilde \mu_1= 1.9 \times 10^{-8} \, , ~~
  \tilde \mu_2 = - 4.7 \times 10^{-6} \, , ~~ \lambda_{231} \sim 10^{-4}, ~~
  \lambda_{123} = -3.2 \times 10^{-4} ~\text{for}~ \tan\beta = 10 \, .
\label{eqn:ToyCouplingValues}
\end{eqnarray}
To illustrate how this coupling pattern can arise from a flavor
symmetry we are providing an exemplary flavor group for this
case. However, since this choice is not necessarily unique and our
conclusions do not depend on the specific flavor group, we omit this
exercise for the other scenarios. In this case, the required relative
suppression can be reproduced by a $Q_X = Z_{4}\times Z_{8}$ family
symmetry with a breaking parameter $\varepsilon$.  The necessary
charge assignments are given by,
\begin{eqnarray}
Q_{B}(L_{1}) = (2,5) \, ,~
Q_{B}(L_{2}) = (0,5) \, ,~
Q_{B}(L_{3}) = (3,2) \, ,
\label{eqn:LCharges}
\end{eqnarray}
which imply 
\begin{eqnarray}
Q_{B}(L_{1}L_{2}E_{3}^{C}) = (3,0) \, ,~
Q_{B}(L_{1}L_{3}E_{2}^{C}) = (1,2) \, ,~
Q_{B}(L_{2}L_{3}E_{1}^{C}) = (1,2) \, .
\label{eqn:LLECharges}
\end{eqnarray}
These assignments lead exactly to the required suppression of the
couplings with $\varepsilon \approx 0.1$ as 
\begin{eqnarray}
\mu_2 (=\mu_{3}) \sim \varepsilon^{5} \, , ~
\mu_1 \sim \varepsilon^{7} \, ,~
\lambda_{123} (=\lambda_{132}) \sim \varepsilon^{3}, \, ~ 
\lambda_{231} \sim  \varepsilon^{3} \, .
	\label{eqn:CouplingSuppression}
\end{eqnarray}
In the above example, the near equality of the magnitude of the
entries in the $\mu-\tau$ block is ensured by saturating them with the
tree level contributions, while keeping the loop contributions
suppressed. If, within the TBM framework, we now consider $m_3$ to be
slightly above zero, then $m_{\mu\mu}=m_{\tau\tau}>|m_{\mu\tau}|$. To
obtain $m_3 = 0.001$ eV with $\tilde m = \mu = 100$ GeV, we need
$\tilde \mu_1 = 1.9 \times 10^{-8}$, $\tilde \mu_2 = - 4.6 \times
10^{-6}$, $\tilde \mu_3 = 4.7 \times 10^{-6}$, $\lambda_{123} = - 3.1
\times 10^{-4}$, $\lambda_{132} = - 3.3 \times 10^{-4}$,
$\lambda_{231} = 2.7 \times 10^{-3}$.  We should note two important
things: ($i$) These choices imply $\lambda_{231} \lambda'_3 = 1.3
\times 10^{-8}$, which mildly overshoots the $K_L \to \mu e$ bound as
shown in Eq.~(\ref{prod-2}). If we increase $m_3$ further, the
disagreement with the $K_L$ bounds deepens. ($ii$) The `four
parameter' scenario with $|\mu_2| = |\mu_3|$ and $|\lambda_{123} =
\lambda_{132}|$ is not compatible with a non-vanishing absolute
neutrino mass scale, i.e. we cannot fit the data assuming these
`equalities' with $m_3>0$, because of the hierarchical nature of the
charged lepton masses which appear in Eq.~(\ref{eqn:nmmDominant}).

{\em Normal hierarchy:}~ We now consider normal hierarchy of neutrino
masses. In this case the smallest mass eigenvalue is $m_{1}$. Within
the TBM structure if we keep $m_{1}=0$, it follows that
$m_{\mu\mu}=m_{\tau\tau}>|m_{\mu\tau}|$.  The numerical values of the
mass matrix entries are
\begin{align}
m = \begin{pmatrix}
2.90\times 10^{-3} & 2.90\times 10^{-3} & -2.90\times 10^{-3}	\\
2.90\times 10^{-3} & 2.80 \times 10^{-2} & 2.21 \times 10^{-2}	\\
-2.90\times 10^{-3} & 2.21 \times 10^{-2} & 2.80 \times 10^{-2}	
\end{pmatrix}\,\text{eV}\ \text{(NH, TBM, $m_1=0$)}.
\label{nh-tbm-m30}
\end{align}
The couplings needed to fit these entries are $\tilde \mu_1 = -5.2
\times 10^{-7}$, $\tilde \mu_2 = 3.9 \times 10^{-6}$, $\tilde \mu_3 =
5.0 \times 10^{-6}$, $\lambda_{123} = - 4.4 \times 10^{-3}$,
$\lambda_{132} = - 1.2 \times 10^{-6}$, $\lambda_{231} = 1.0 \times
10^{-3}$. Although we are within the $K_L \to \mu e$ bound, the
requirement $m_{e\mu} = - m_{e\tau}$ is realized quite
differently. The relative signs of the tree-level couplings invariably
imply $m_{e\mu}^{\text{tree}} \approx +m_{e\tau}^{\text{tree}}$. This
difference between the experimental requirement and the tree-level
contribution cannot be resolved, even keeping in mind that signs of
the RpV couplings can be chosen at will and also each neutrino field
can be redefined to absorb a sign. Therefore, a sign adjustment for
one of the entries ($e \mu$) via a large loop contribution is needed,
while the loop contribution to the other one ($e \tau$) becomes
negligible. This is reflected in the large hierarchy between
$\lambda_{123}$ and $\lambda_{132}$. We recall that such a sign
adjustment was not required in the case of inverted hierarchy (TBM,
$m_3 = 0$). If we now increase the value of $m_1$ (from zero) and try
to fit normal hierarchy within the TBM framework, the $K_L \to \mu e$
bound haunts us like in the case of inverted hierarchy with $m_3 >
0$. Therefore, our most robust prediction is the tight constraint for
the smallest mass eigenvalue. Thus, inverted hierarchy can be fit with
four parameters, while normal hierarchy requires six parameters and a
sign altering large loop correction.

\section{Non-zero $\theta_{13}$}
In view of the recent T2K data which measures non-vanishing
$\theta_{13}$, we study how flexible we are to accommodate normal or
inverted hierarchies while keeping $\theta_{13}$ close to its central
value of $9.0^\circ$.  Unlike in the case of TBM which guarantees
$\left|m_{e\mu}\right| = \left|m_{e\tau}\right|$ and $m_{\mu\mu} =
m_{\tau\tau}$, it is not possible to fit the data with 4 parameters
when $\theta_{13} \ne 0$.

{\em Inverted hierarchy:}~ First we consider the case $m_3 = 0$. The
numerical entries of the mass matrix are given by 
\begin{align}
m = \begin{pmatrix}
4.80\times 10^{-2} & -5.13 \times 10^{-3} & -5.63 \times 10^{-3}	\\
-5.13 \times 10^{-3} & 2.53 \times 10^{-2} & -2.41 \times 10^{-2} \\
-5.63 \times 10^{-3} & -2.41 \times 10^{-2} & 2.54 \times 10^{-2}	
\end{pmatrix}\,\text{eV}\ \text{(IH, $\theta_{13}= 9.0^\circ$, $m_3=0$)}
\, .
\label{thneq0}
\end{align}
This can be fit with $\tilde \mu_1 = -1.1 \times 10^{-6}$, $\tilde
\mu_2 = -4.5 \times 10^{-6}$, $\tilde \mu_3 = 4.8 \times 10^{-6}$,
$\lambda_{123} = 9.3 \times 10^{-3}$, $\lambda_{132} = 1.1 \times
10^{-5}$, $\lambda_{231} = -1.1 \times 10^{-4}$. Two things are worth
noting: ($i$) The magnitudes of $\lambda_{123}$ and $\lambda_{132}$
are separated by nearly three orders, while they assumed identical
numerical values in the case of TBM. ($ii$) The tree-level
contribution to $m_{e\mu}$ has the wrong sign like in the case of NH
with $\theta_{13}$=0. Again a large sign adjusting loop contribution
is needed to be in agreement with the experimental data. If we now
increase the value of $m_3$, the required magnitude for
$\lambda_{231}$ becomes larger, and eventually beyond $m_3 = 0.01$ eV
the $K_L \to \mu e$ bound overshoots.

{\em Normal hierarchy:}~   For $m_1 = 0$, the mass matrix entries are
given by 
\begin{align}
m = \begin{pmatrix}
4.06\times 10^{-3} & 8.02 \times 10^{-3} & 2.29 \times 10^{-3}	\\
8.02 \times 10^{-3} & 2.67 \times 10^{-2} & 2.16 \times 10^{-2} \\
2.29 \times 10^{-3} & 2.16 \times 10^{-2} & 2.80 \times 10^{-2}	
\end{pmatrix}\,\text{eV}\ \text{(NH, $\theta_{13}= 9.0^\circ$, $m_1=0$)}
\, .
\label{thneq1}
\end{align}
This can be reproduced with $\tilde \mu_1 = 4.1 \times 10^{-7}$,
$\tilde \mu_2 = 3.8 \times 10^{-6}$, $\tilde \mu_3 = 5.0 \times
10^{-6}$, $\lambda_{123} = -5.3 \times 10^{-3}$, $\lambda_{132} = -1.5
\times 10^{-6}$, $\lambda_{231} = 8.3 \times 10^{-4}$.  Note that
$\lambda_{231}$ is small enough to satisfy the $K_L \to \mu e$ bound.
Contrary to the case of inverted hierarchy, now no large sign-flipping
correction for $m_{e\mu}$ is needed. However, the difference between
the values of $m_{e\mu}$ and $m_{e\tau}$ still leads to a hierarchy in
the $\lambda$-couplings. Just like in the previous cases, the $K_L \to
\mu e$ bound begins to be relevant as soon as $m_1$ increases to
around 0.005 eV (which requires $\lambda_{231} = 1.5 \times
10^{-3}$). The main conclusion for non-zero $\theta_{13}$ is again
that the smallest mass eigenvalue is required to be almost vanishing
in both hierarchies. But contrary to the TBM case, now IH requires a
sign adjustment, while NH does not.

\section{Collider signatures}	
The LHC signatures of the $\lambda_{ijk}$ couplings have recently been
explored in \cite{Bomark:2011ye}. In our scenario, only three
couplings $\lambda_{ijk} (i\neq j \neq k)$ are relatively large
($10^{-3}-10^{-4}$), the rest are of order $10^{-6}$.  The large
couplings are small enough to make sure that the RpV vertex is
numerically relevant only at the end of a supersymmetry cascade when
the lightest neutralino decays via a $\lambda_{ijk}$ interaction,
$\tilde{\chi}_{1}^{0}\rightarrow l^{\pm}l^{\mp}\nu$. The
$\lambda_{ijk}$ couplings thus give rise to $l_{i}l_{k}$ or
$l_{j}l_{k}$ final states plus missing energy. Depending on the
numerical values of of the corresponding $\lambda_{ijk}$ couplings the
branching ratios into the $l_{i}l_{k}$ or $l_{j}l_{k}$ channel will
scale as $|\lambda_{ijk}|^{2}$. Thus both invariant mass distributions
and number counting of the final state leptons should be a part of the
search method.  However, other decay channels like
$\tilde{\chi}_{1}^{0}\rightarrow W^{\pm}l^{\mp}$ and
$\tilde{\chi}_{1}^{0}\rightarrow Z\nu$ are available due to the
presence of the bilinear couplings. Their role has been investigated
in detail in \cite{Porod:2000hv}. Therefore, a detailed study of
neutralino decays is important to test this and other RpV models and
differentiate between them. The non-observation of an excess in four
lepton events at CMS and ATLAS so far indicates a somewhat heavier
squark mass scale than the one we choose. However, scaling the slepton
masses accordingly, this will not lead to any significant changes
related to our work.

\section{Conclusions}	

In this paper we have studied a generic and simple flavor model which
reduces the number of independent couplings from 39 to 6,
i.e. $\mu_{i}\,(i=1,2,3)$, $\lambda_{123}$, $\lambda_{132}$ and
$\lambda_{231}$.  This results in an extremely predictive framework,
which can reproduce the correct neutrino masses and mixings while
satisfying all other low energy bounds.

In its simplest realization the scenario leads to a four parameter
model with exact tri-bimaximal mixing and prefers inverse hierarchy,
for which a specific flavor model, viz $Z_{4}\times Z_{8}$, has been
proposed. A non-vanishing mixing angle $\theta_{13}$ can be
accommodated in a six parameter realization.

A general prediction of all possible realizations is an almost
vanishing absolute mass scale for neutrinos, i.e. an essentially
massless lightest neutrino. This feature is tightly related to the
non-observation of $K_{L}\rightarrow \mu e$ which affects many
important coupling products in this framework.  As a consequence, any
positive signal in one of the upcoming neutrinoless double beta decay
experiments would imply an inverted neutrino mass hierarchy, since for
the combination of normal hierarchy and an almost vanishing absolute
mass scale, the resulting $|m_{ee}|$ is beyond their sensitivity. In
other words, if a conclusive evidence of nonzero $\theta_{13}$ is
established, then our scenario would be able to accommodate a positive
signal of neutrinoless double beta decay only at the expense of large
sign-flipping correction to one of the off-diagonal elements of the
mass matrix.  Moreover, the flavor structure proposed here can lead to
specific decays of a neutralino LSP at the LHC.

\section*{Acknowledgments:} 
We thank W.Porod for valuable suggestions.  This work was supported by
DAAD-DST PPP Grant No.~D/08/04933 and DST-DAAD project
No.~INT/DAAD/P-181/2008.  G.B. acknowledges the {\em Gambrinus} grant
and hospitality at T.U. Dortmund during a part of the
collaboration. D.P. and H.P. were supported by DFG Grant No. PA
803/5-1, the Physics at the Terascale Helmholtz Alliance Working
Group: Neutrino masses and Lepton Flavor Violation at the LHC, and
they acknowledge hospitality at the Saha Institute of Nuclear Physics,
Kolkata, during a part of this collaboration.

\end{document}